\documentclass[iop]{emulateapj}
\usepackage{apjfonts}
\usepackage{times,graphicx,aas_macros}
\usepackage{amsmath}
\usepackage{float}
\usepackage{ifthen}
\usepackage{times}
\usepackage{natbib}
\usepackage{rotating}

\begin{document}

\shorttitle{Polarized Ly$\alpha$ at z$=$2.3}
\shortauthors{Andrew Humphrey {\it et al.}}

\title{Polarized extended Ly$\alpha$ emission from a z$=$2.3 radio galaxy}

\author{A.~Humphrey\altaffilmark{1}, J.~Vernet\altaffilmark{2},
M.~Villar-Mart\'{i}n\altaffilmark{3}, S.~di Serego Alighieri\altaffilmark{4}, 
R.A.E.~Fosbury\altaffilmark{2}, A.~Cimatti\altaffilmark{5} }

\altaffiltext{1}{Centro de Astrof\'{i}sica da Universidade do Porto, Rua das Estrelas, 4150-762, Porto, Portugal. Email: andrew.humphrey@astro.up.pt}
\altaffiltext{2}{European Southern Observatory, Karl-Schwarzschild-Strasse 2, D-85748 Garching, Germany}
\altaffiltext{3}{Centro de Astrobiolog\'{i}a (INTA-CSIC), Carretera de Ajalvir, km 4, 28850 Torrej\'on de Ardoz, Madrid, Spain}
\altaffiltext{4}{INAF-Osservatorio Astrofisico di Arcetri, L.go E. Fermi 5, 50125 Firenze, Italy}
\altaffiltext{5}{Dipartimento di Astronomia, Universit\`a di Bologna, Via Ranzani 1, I-40127 Bologna, Italy}

\begin{abstract}
We present spatially resolved spectropolarimetic measurements of the 100-kpc scale gaseous environment of the z$=$2.34 radio galaxy TXS 0211-122.  The polarization level of the narrow Ly$\alpha$ emission is low centrally (P$<$5 \%), but rises to P$=$16.4$\pm$4.6 \% in the Eastern part of the nebula, indicating that the nebula is at least partly powered by the scattering of Ly$\alpha$ photons by HI.  Not only is this the first detection of polarized Ly$\alpha$ around a radio-loud active galaxy, it is also the second detection to date for any kind of Ly$\alpha$ nebula.  We also detect a pair of diametrically opposed UV continuum sources along the slit, at the outer edges of the Ly$\alpha$ nebula, which we suggest may be the limb of a dusty shell, related to the large-scale HI absorbers often associated with high-z radio galaxies.  
\end{abstract}


\section{Introduction}
Luminous high-redshift Ly$\alpha$ nebulae promise to yield significant insights into the physics of massive galaxy formation (e.g., Mori \& Umemura et al. 2006).  As prodigious sources of HI Ly$\alpha$ $\lambda$1216 photons, they provide an efficient way to select distant galaxies expected to be undergoing
 phases of significant mass-assembly (e.g. Steidel et al. 2000).  

What powers the Ly$\alpha$ emission from these nebulae?  This question is fundamental to our understanding of the nature and evolutionary status of these enigmatic objects and their central galaxies.  A number of ideas have been discussed in the literature, the most promising of which are: (i) cooling radiation emitted during the infall of gas into the dark matter halos of massive proto-galaxies (e.g. Haiman, Spanns \& Quataert 2000); (ii) mechanical energy injected by supernovae during powerful starbursts (e.g. Taniguchi \& Shioya 2000); (iii) scattering of Ly$\alpha$ photons by neutral hydrogen halos around
massive, Ly$\alpha$ emitting galaxies (e.g. Villar-Mart\'{i}n et al. 1996; Zheng et al. 2011); or
(iv) photoionization by luminous active galactic nuclei and/or young stars (AGNs;
e.g. Villar-Mart\'{i}n et al. 2003; Prescott et al. 2012).  While there is currently no consensus concerning which
 powering mechanism is dominant, it is now clear that no single mechanism is able to 
explain all Ly$\alpha$ nebulae (e.g. Smith \& Jarvis 2007; Prescott et al. 2009, 2012).  

Scattering of Ly$\alpha$ photons by neutral hydrogen can result in a net observed polarization in the Ly$\alpha$ emission.  Polarimetry is thus a promising way to distinguish between scattering and {\it in situ} production of the Ly$\alpha$ photons (e.g. Dijkstra \& Loeb 2008). Due to the challenges and observational expense involved, polarization measurements of spatially extended Ly$\alpha$ emission had been, until now, attempted for only two Ly$\alpha$ nebulae.  Prescott et al. (2011) presented imaging polarimetry of the Ly$\alpha$ emission from a giant Ly$\alpha$ nebula associated with a radio-quiet active galaxy at z$=$2.66, but did not detect significant polarization.  Hayes et al. (2011) targeted a large Ly$\alpha$ nebula in a proto-cluster region at z$=$3.09, again using narrow-band imaging polarimetry, and detected polarized Ly$\alpha$: They measured P=11.9$\pm$2\% within a radius of 7\arcsec, with a trend for P to be higher at larger radii, or in regions of low Ly$\alpha$ surface-brightness, and with electric field vectors consistent with illumination by a central source.  A clear general picture of the polarization properties of high-redshift Ly$\alpha$ nebulae has not yet emerged, and more nebulae, with a wider range of properties, now need to be tested.  

The resonant nature of Ly$\alpha$ also allows detection of associated gaseous structures in front of the Ly$\alpha$ emitting source.  Many giant Ly$\alpha$ nebulae do indeed show strong HI absorption features in their Ly$\alpha$ velocity profile (e.g. van Ojik et al. 1997; Fosbury et al. 2003; Wilman et al. 2005; Humphrey et al. 2008a).  These HI absorption features are usually black at their center, usually detected across the full spatial extent of the Ly$\alpha$ emission, and usually show a blueshift from the center of the Ly$\alpha$ emission (see van Ojik et al. 1997), implying that the absorber is an expanding shell surrounding the Ly$\alpha$ nebula (Binette et al. 2000).  The minimum shell radius, typically about $\ga$50 kpc, allows important constraints on other properties of the shells (mass, kinetic energy, etc.; e.g. van Ojik et al. 1997; Binette et al. 2006; Humphrey et al. 2013).  Detecting these absorbing structures in {\it emission} would substantially improve our understanding of their physical properties, their spatial location, and their origins.  

\begin{figure*}
\includegraphics{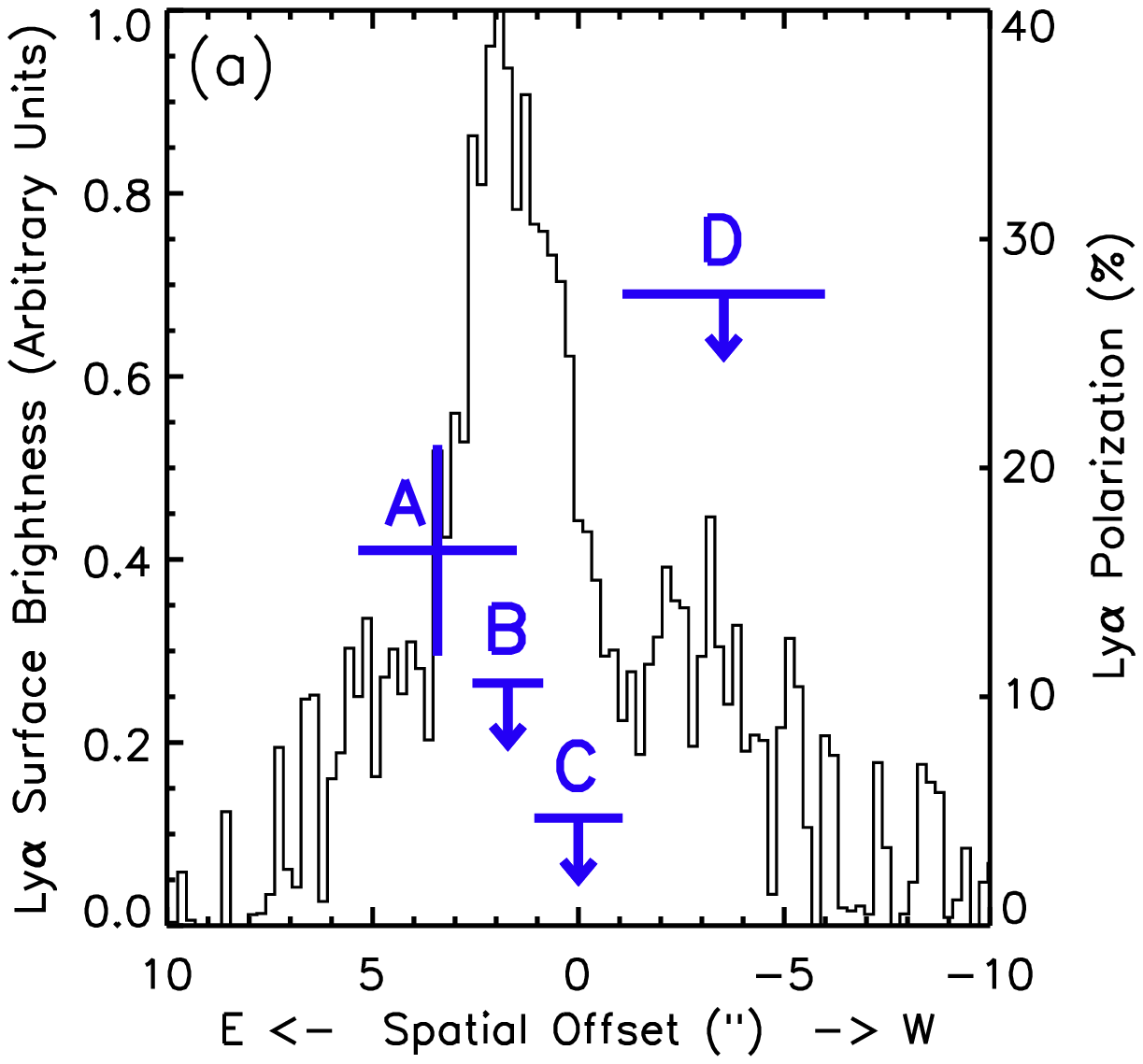}
\includegraphics{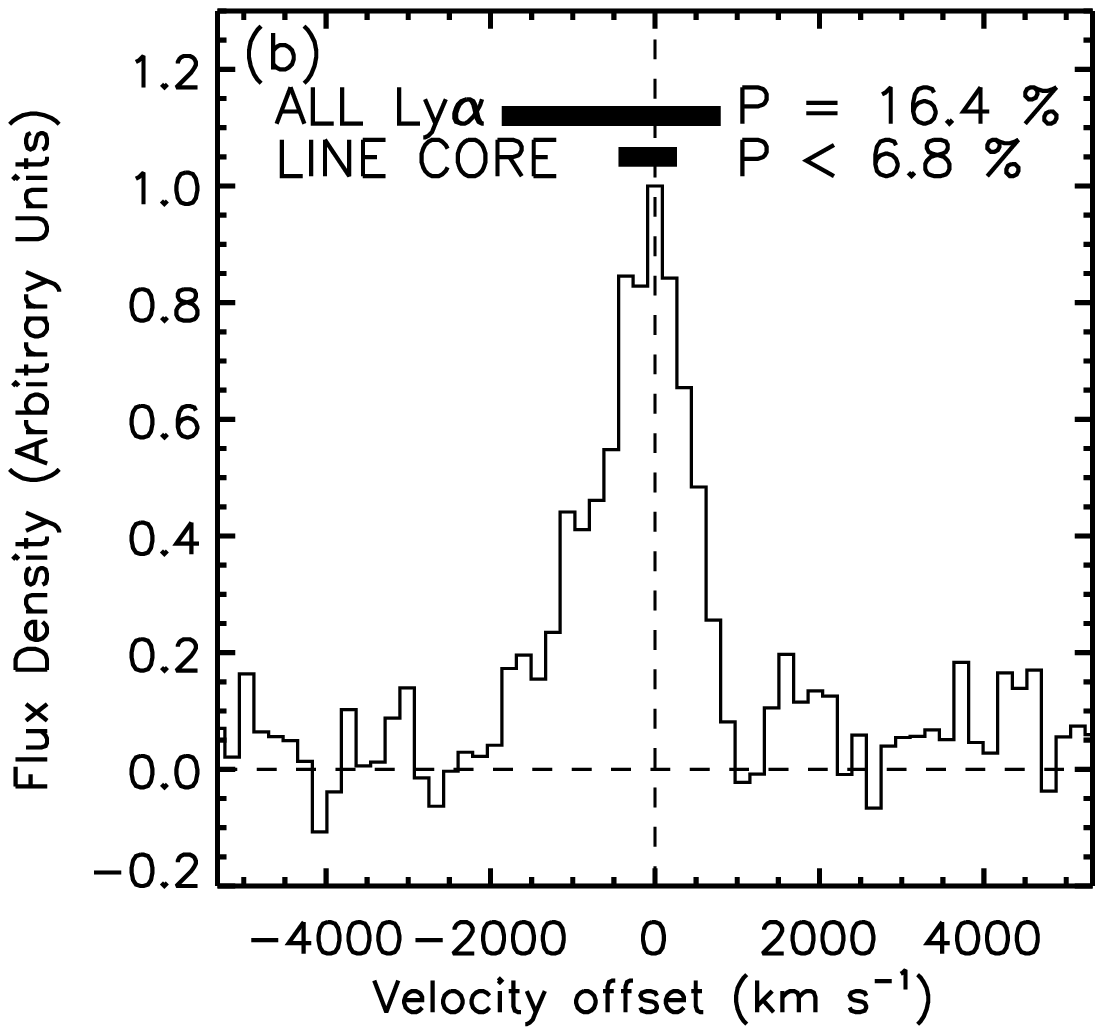}
\vspace{3.2in}
\caption{(a) Spatial profile of continuum-subtracted Ly$\alpha$ emission along the slit (continuous curve; left axis), with the Ly$\alpha$ polarization measurements (points A-D; right axis; aperture definition given in Table 1).  The spatial zero is the peak of the UV continuum.  (b) The spectral region around Ly$\alpha$ in Aperture A, with Ly$\alpha$ polarization measurements in two spectral apertures.}
\end{figure*}

In this Letter, we analyze deep spectroscopy and spectropolarimetry of TXS 0211-122 -- a z$=$2.34 radio galaxy known to be associated with a large-kpc scale Ly$\alpha$ nebula, and with a $\ge$100 kpc scale HI absorbing structure (van Ojik et al. 1994, 1997).  We aim to probe the geometry, powering mechanism, and
 origin of the nebular and absorbing structures in this galaxy's gaseous environment; understanding the physics of these systems is also important to
 shed light on the relevant feedback processes in the formation and evolution of massive galaxies.  We assume $H_{0}$=71 km s$^{-1}$ Mpc$^{-1}$, $\Omega_{\Lambda}$=0.73 and $\Omega_{m}$=0.27, with 1\arcsec corresponding to 8.30 kpc at z=2.34.

\section{Observations}
TXS 0211-122 was observed using the Low Resolution Imaging Spectrometer (Oke et al. 1995) in polarimetry mode (Goodrich et al. 1995), in sub-arcsecond seeing at the Keck II telescope, during 1997 December 24-26.  The 300 line mm$^{-1}$ grating and 1\arcsec~ wide slit resulted in an instrumental profile (IP) of 10 \AA, a dispersion of 2.4 \AA~ per pixel, and a spectral range of $\sim$3900-9000 \AA~ ($\sim$1200-2700 \AA~ in the rest-frame).  The 24\arcsec~ long slit was oriented approximately along the major axis of the radio source (PA=104$^{\circ}$, North through East), to maximize signal to noise for extended emission.  The total on-object exposure time was 28580 s, split into four sets of integrations, each with four integrations with the half-wave plate successively at PA=0$^{\circ}$, 45$^{\circ}$, 22.5$^{\circ}$, 67.5$^{\circ}$.  Full details are given by Vernet et al. (2001: V01 hereinafter).  

The target was also observed using ISAAC (Moorwood et al. 1998) in long-slit mode at the Very Large Telescope, on 1999 November 27-28.  The lines [OII] $\lambda$3727, [OIII] $\lambda\lambda$4959,5007 and H$\alpha$ lie in relatively transparent regions of the J, H and K-bands, and integration times were 5200 s, 7200 s and 7200 s, respectively.  The low resolution grating was used with a 1\arcsec~ slit to give an IP of $\sim$30 \AA~ in J and H-bands, and $\sim$50 \AA~ in the K-band.  The slit PA was also 104$^{\circ}$.  Complete details of the ISAAC data are given by Humphrey et al. (2008b).  

\section{A polarized Ly$\alpha$ nebula}
As previously discussed by van Ojik et al. (1994) and Villar-Mart\'{i}n et al. (2003), TXS 0211-122 is associated with an extended Ly$\alpha$, CIV $\lambda$1549 and HeII $\lambda$1640 nebula, which shows a total Ly$\alpha$ extent along the slit of $\sim$13\arcsec, or $\sim$110 kpc.  Its HeII velocity curve displays kinematics that overall are very quiescent (Villar-Mart\'{i}n et al. 2003), consistent with infall (Humphrey et al. 2007), although near the galactic nucleus an excess of emission in the blue wing of HeII suggests the presence of outflowing gas in addition (Humphrey et al. 2006).  At least in the bright, central few tens of kpc, the nebula is ionized predominantly by the radiation field of the active galactic nucleus, and has approximately solar metallicity (V01; Humphrey et al. 2008b).  The spatial distribution of the Ly$\alpha$ emission along the slit is strikingly asymmetrical compared to the continuum and the other emission lines (Fig. 1 and 2), suggesting that the Ly$\alpha$ emission is either strongly absorbed, or is powered
 by a different mechanism than the other lines (see also van Ojik et al. 1994; Villar-Mart\'{i}n et al. 2003; Humphrey et al. 2007).  

The polarization of the Ly$\alpha$ emission line can provide useful constraints on which excitation process powers the giant Ly$\alpha$ nebula (e.g. Dijkstra \& Loeb 2008; Dijkstra \& Kramer 2012).  To this end, we have taken polarization measurements of the Ly$\alpha$ emission at different spatial positions along the slit.  Wavelength intervals were defined as the contiguous range of dispersion pixels with Ly$\alpha$ detected at S/N$\ge$2 pixel$^{-1}$.  Sky-subtraction was performed using a line and continuum free spatial bin covering an identical spectral range.  Unbiased values of the percentage polarization were estimated according to Simmons \& Stewart (1985), and uncertainties were determined using a Monte-Carlo method considering detector noise and background polarization (Vernet 2000).  Throughout this analysis we adopt the spatial peak of the UV continuum (and H$\alpha$ emission) along the slit as our fiducial position of the galactic nucleus of the radio galaxy.  Table 1 and Figure 1 give the results of our Ly$\alpha$ polarization measurements.  All upper or lower limits quoted in this Letter are 3$\sigma$.  

\begin{figure*}
\includegraphics{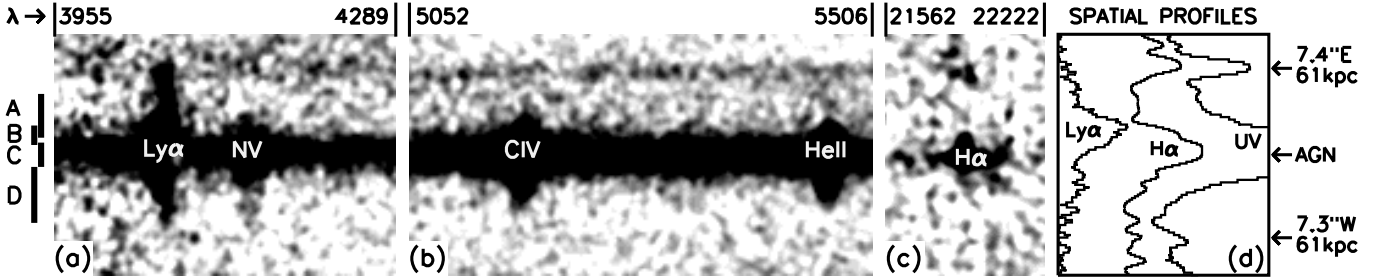}
\vspace{1.5in}
\caption{Distribution of line and continuum emission along the slit.  (a)-(c) show cut-outs from the two-dimensional spectra, smoothed by an FWHM$=$0.7\arcsec Gaussian.  The spatial axis (vertical) and is centered on the UV continuum peak (AGN); the dispersion axis (horizontal) is discontinuous, with wavelength ranges (\AA) given above the panels.  Major emission lines are labeled in white. (d) shows spatial profiles of continuum-subtracted Ly$\alpha$ and H$\alpha$ emission, and the rest-frame UV continuum between CIV and HeII.  The Ly$\alpha$ distribution differs very slightly between (a) and (d) due to continuum subtraction in the latter.}
\end{figure*}

We detect polarized Ly$\alpha$ emission, with P$=$16.4$\pm$4.6 \% (S/N$=$3.6) and a PA of 154$\pm$8$^{\circ}$, in a region 1.50\arcsec-5.35\arcsec~ East of the nucleus (aperture A in Table 1 and Figure 1).  The wavelength range covered by this
 aperture is 4046-4081 \AA,  and its Ly$\alpha$ emission has a total-flux S/N of 32.  Weak continuum emission is also present at this position, with P$=$15.3$\pm$3.3 \% and PA$=$194$\pm$6$^{\circ}$ measured in a wide bin covering the relatively line-free continuum region between the wavelengths of the NV and CIV lines.  Thus, we have repeated the Ly$\alpha$ polarization measurement after subtracting the continuum emission that lies immediately 
 redward of Ly$\alpha$ (and NV) at 4090-4140 \AA, resulting in the noisier but nonetheless consistent measurement P$=$23.1$\pm$7.1 \% with
 PA$=$143$\pm$8$^{\circ}$.  The core of Ly$\alpha$ emission (4067-4074 \AA), however, does not show significant polarization at this spatial position
 (P$\le$6.8 \%).  In the same spatial aperture, the emission lines CIV and HeII have P$\le$10 and P$\le$13 \%, respectively.  The high surface-brightness Ly$\alpha$ emission closer to the nucleus is also not significantly polarized
 (apertures B and C: $\le$10.6 and $\le$4.7 \%), while in the Western nebula the data do not provide the necessary sensitivity for useful constraints (aperture D: P$\le$27.6 \%).  

As a check for our Ly$\alpha$ polarization detection in aperture A, Table 2 lists Ly$\alpha$ polarization measurements using different combinations of the four integration sets.  Although noisier, these measurements show generally good consistency.  

This is the first detection of polarized narrow Ly$\alpha$ emission from a nebula associated with a radio-loud active galaxy, and only the second from an extragalactic nebula of any kind (Hayes et al. 2011; see also Prescott et al. 2011).  Our measurement of P$=$16.4$\pm$4.6 \% for Ly$\alpha$ is 
 consistent with the range in polarization predicted by models of Ly$\alpha$ scattering in HI halos around high-z galaxies (up to $\sim$40 \%: Dijkstra \&
 Loeb 2008; Dijkstra \& Kramer 2012).  Together with the recent detection of polarized narrow Ly$\alpha$ in a high-z Ly$\alpha$ blob by Hayes et al. (2011), our new detection in TXS 0211-122 then suggests it may not be uncommon for Ly$\alpha$ nebulae at high redshift to be powered (at least partly) by scattering of Ly$\alpha$ by HI.  Polarization measurements of a larger sample are now needed to ascertain whether or not this is generally the case for high-z Ly$\alpha$ nebulae.  

Interestingly, the Ly$\alpha$ polarization PA (154$\pm$8$^{\circ}$) differs from that of the UV continuum by $\sim$30$^{\circ}$ (Table 3; also V01), implying that the resonantly scattered Ly$\alpha$ photons and dust-scattered UV continuum photons have different scattering geometries.

\section{A 120 kpc gaseous shell around the Ly$\alpha$ nebula?}
At the Eastern edge of the Ly$\alpha$ nebula we detect a faint, unresolved 
continuum source (FWHM$=$1.1$\pm$0.1\arcsec), positioned 7.4$\pm$0.2\arcsec~ E (61$\pm$2 kpc)
along the slit, and a fainter one
diametrically opposed at a position 7.3$\pm$0.3\arcsec~ W (61$\pm$2
kpc; Figure 2).  
The Eastern source is polarized, with P$=$22$\pm$6
\% and PA$=$182$\pm$7$^{\circ}$ measured in the line-free region redward of the 
wavelength of the NV emission line, at 4260-4600 \AA~ (or 1280-1380 \AA~ in the restframe).  
This polarization PA is close to perpendicular to the slit and to the radio source.  For comparison, 
we measure P$=$21$\pm$2 \% and PA$=$185$\pm$2$^{\circ}$ in the same wavelength range using
 a 4.1\arcsec aperture centered on the position of the central radio galaxy (see also V01).
Like the radio galaxy, the Eastern source shows a strong gradient (or decrement) in P towards 
longer wavelengths, falling to P$\le$10 \% in the continuum region between the expected wavelengths of 
CIV and HeII.  This wavelength dependence is consistent with a mixture of nebular continuum, which drops 
off sharply at $\lambda$$\la$1300 \AA, and scattered light (V01).  

The Eastern source is also detected in
H$\alpha$ (Figure 2), with FWHM$=$1000$\pm$300 km s$^{-1}$,
and z$=$2.335$\pm$0.004 confirming its association with the
radio galaxy.  Such a high FWHM suggests turbulent, rather than gravitational, motion.
Table 3 provides properties of this source,
 alongside those of the radio galaxy.  

Previous modeling of the rest-frame ultraviolet 
emission from the extended narrow line region of the radio galaxy has shown 
that it is powered predominantly by illumination by the central AGN 
(V01; Humphrey et al. 2008b).  Interestingly, the spectral properties of the 
Eastern source are very similar to those of the radio galaxy, in terms of the percentage and
wavelength dependence of polarization, and also the relative fluxes of Ly$\alpha$, 
H$\alpha$ and the UV continuum (Table 3).  This, together with the fact that the polarization PA is 
perpendicular to the slit (and to the radio source), leads us to conclude that the Eastern source 
is also powered by illumination by the AGN.  

The Western UV source is detected across the full wavelength range between Ly$\alpha$ and CIII] $\lambda$1908, outside of 
which the sensitivity of the spectrum rapidly declines.  Its continuum flux density is 
0.6$\pm$0.2 $\times$10$^{-19}$ erg s$^{-1}$ cm$^{-2}$ \AA$^{-1}$ at 5300 \AA. 
Unfortunately, we cannot ascertain its redshift because we detect no lines therefrom.  

A natural explanation for the diametric opposition of the two UV sources
is that they are the limb of a giant shell of gas and dust centered on the radio galaxy, 
detected in emission thanks to illumination by the active nucleus.  
Provided it has a sufficiently high covering factor, a shell like this also ought to be detected via strong, spatially extended absorption lines,
 including Ly$\alpha$ (Humphrey et al. 2013).  Indeed, TXS 0211-122 does show a strong, spatially extended Ly$\alpha$ absorption feature in its spectrum when observed at moderate 
spectral resolution (van Ojk et al. 1997), as do around half of all radio-loud galaxies at high redshift (e.g. van Ojk et al. 1997; Binette et al. 2000; Baker et al. 2002; Humphrey et al. 2008a).  These giant HI absorbers are usually interpreted 
in terms of a giant shell of gas enveloping the galaxy and its Ly$\alpha$ nebula (Binette et al. 2000).  
More specifically, TXS 0211-122 shows an HI absorber with a blueshift of $\sim$60 km s$^{-1}$ from the Ly$\alpha$ emission, 
 an observed diameter of $\ge$10\arcsec~ ($\ge$83 kpc) in the plane of the sky (comparable to the size of the Ly$\alpha$ emitting nebula), 
a hydrogen column density of $N_{H}\ga$10$^{18}$ cm$^{-2}$, and covering factor approaching unity (van Ojik et al. 1997).  

We suggest that the diametrically opposed UV sources and the HI absorption are manifestations of a single structure: a shell of gas and 
dust enveloping the TXS 0211-122 and its Ly$\alpha$ nebula.  If this shell has a radius of 61 kpc and a hydrogen column density of 
$N_{H}\ga$10$^{18}$ cm$^{-2}$, its total gas mass would then be $\ga$4$\times$10$^{8}$ $M_{\odot}$.  

Although a pair of companion galaxies could be a potential alternative interpretation of the diametrically opposed UV sources,
 such a scenario would require a very fortuitous alignment between the two UV galaxies and TXS 0211-122.  Deep 
narrow-band imaging or integral
 field spectroscopy would allow a more definitive resolution of this issue.

\begin{table*}
\centering
\caption{Ly$\alpha$ polarization of TXS 0211-122.} 
\begin{tabular}{lllllll}
\hline
Ap. & Position (\arcsec) & $\lambda$-range (\AA) & P (\%) & PA ($^{\circ}$) & Ly$\alpha$/HeII$^a$ & Comments \\
\hline
A & 1.50--5.35 E& 4046--4081& 16.4$\pm$4.6 & 154$\pm$8  & 5.5$\pm$0.8 & East Ly$\alpha$ region \\
A & 1.50--5.35 E& 4067--4074& $\le$6.8 (3$\sigma$) & --  & 5.5$\pm$0.8 & East Ly$\alpha$ region; line core only \\
A & 1.50--5.35 E& 4046--4081& 23.1$\pm$7.1 & 143$\pm$8  & 5.5$\pm$0.8 & East Ly$\alpha$ region; continuum-subtracted \\
B & 0.86--2.57 E& 4046--4081& $<$10.6 (3$\sigma$) & -- & 2.0$\pm$0.2 & At the spatial peak of Ly$\alpha$ \\
C & 1.07 W -- 1.07 E & 4046--4077& $<$4.7 (3$\sigma$) & -- & 0.42$\pm$0.02 & At the spatial peak of UV continuum \\
D & 1.07--5.99 W& 4053--4081& $<$27.6 (3$\sigma$) & -- & 1.3$\pm$0.2 & Western Ly$\alpha$ nebula \\
\hline
\end{tabular}
\raggedright
\\
$^a$ Line ratio Ly$\alpha$ / HeII $\lambda$1640, determined from the total line fluxes at the given spatial positions along the slit.
\end{table*}

\begin{table}
\centering
\caption{Ly$\alpha$ polarization in aperture A for different combinations of the four integration sets.} 
\begin{tabular}{lll}
\hline
Run & P(\%) & PA ($^{\circ}$) \\
\hline
1+2 & 16.6$\pm$6.3 & 182$\pm$10 \\
1+3 & 25.4$\pm$7.6 & 167$\pm$8 \\
1+4 & 10.2$\pm$6.8 & 161$\pm$18 \\
2+3 & 26.2$\pm$9.4 & 147$\pm$8 \\
2+4 & 14.3$\pm$7.3 & 126$\pm$14 \\
3+4 & 35.7$\pm$10.0 & 137$\pm$6 \\
1+2+3 & 19.2$\pm$5.4 & 166$\pm$8 \\
1+2+4 & 9.1$\pm$5.2 & 161$\pm$17 \\
1+3+4 & 19.9$\pm$5.7 & 154$\pm$8 \\
2+3+4 & 23.9$\pm$6.4 & 139$\pm$7 \\
1+2+3+4 & 16.4$\pm$4.6 & 154$\pm$8 \\
\hline
\end{tabular}
\end{table}

\begin{table}
\centering
\caption{Properties of the Eastern continuum source and TXS 0211-122.} 
\begin{tabular}{llll}
\hline
Property & $\lambda$-interval & Eastern source & TXS 0211-122$^a$ \\
\hline
Offset & 4000--7000 & 7.4$\pm$0.2\arcsec~E & 0.0\arcsec \\
$F_{\lambda}$ & 4200--4300 & 5.1$\pm$0.4$^b$ & 23.0$\pm$0.7$^b$ \\
$F_{\lambda}$ & 5250--5400 & 2.5$\pm$0.2$^b$ & 24.7$\pm$0.3$^b$ \\
$F_{\lambda}$ & 6100--6290 & 3.3$\pm$0.2$^b$ & 20.1$\pm$0.2$^b$ \\
Pol UV$^c$ & 4260--4600 & 22$\pm$6\% 182$\pm$7$^{\circ}$ & 21$\pm$2\% 185$\pm$2$^{\circ}$ \\
Pol UV$^c$ & 5240--5430 & $\le$10\% & 18$\pm$1\% 188$\pm$2$^{\circ}$ \\
$F$ & Ly$\alpha$ & 1.4$\pm$0.2$^d$ & 9.0$\pm$0.3$^d$ \\
$F$ & CIV & $\le$0.4$^d$ & 28$\pm$1$^d$ \\
$F$ & HeII & $\le$0.3$^d$ & 15.7$\pm$0.3$^d$ \\
$F$ & H$\alpha$ & 7$\pm$2$^d$ & 64$\pm$5$^d$ \\
$F$ & [OIII] 5007 & $\le$8$^d$ & 209$\pm$4$^d$ \\
$F$ & [OII] 3727 & $\le$11$^d$ & 25$\pm$5$^d$ \\
Flux Ratio & Ly$\alpha$/HeII & $\ge$4.7 & 0.57$\pm$0.01 \\
Flux Ratio & Ly$\alpha$/H$\alpha$ & 0.20$\pm$0.06 & 0.14$\pm$0.02 \\
Flux Ratio & HeII/H$\alpha$ & $\le$0.04 & 0.25$\pm$0.02 \\
Flux Ratio & [OIII]/H$\alpha$ & $\le$1.1 & 3.27$\pm$0.02 \\
\hline
\end{tabular}
\raggedright
\\
$^a$ Measured in an aperture of diameter 4.1\arcsec, as also used by Vernet et al. (2001) and Humphrey et al. (2008b)\\
$^b$ $10^{-19}$ erg s$^{-1}$ cm$^{-2}$ \AA$^{-1}$.  \\
$^c$ Polarization (\%) and PA ($^{\circ}$) of the UV continuum measured in these observer-frame wavelength bin.\\
$^d$ $10^{-17}$ erg s$^{-1}$ cm$^{-2}$.
\vspace{0.4in}
\end{table}

\section{Discussion: The origin of the Ly$\alpha$ nebula and shell}
We now propose a scenario for the origin of the Ly$\alpha$ nebula and the gaseous shell associated with TXS 0211-122.  
The presence of dust and metals in the shell argue strongly for an internal origin for this structure, rather than an external one.  
We suggest that this shell is the result of a starburst event in the central galaxy, during which mechanical energy is deposited into the 
interstellar medium by supernova explosions, to produce expanding bubbles of gas.  Expanding bubbles produced by the individual supernovae sweep up material from the interstellar medium, merging with other bubbles as they expand, to eventually form a single superbubble of gas and dust centered on the galaxy.  In the presence of a sufficiently luminous background source (e.g., the central radio galaxy), this bubble will manifest itself as a spatially extended, narrow Ly$\alpha$ absorption feature with a blueshift with respect to the velocity of the Ly$\alpha$ {\it emission} (e.g. Tenorio-Tagle et al. 1999; Binette et al. 2000).  Moreover, depending on its properties, such as covering factor, radius, column density, etc., the shell may also be detectable in emission where it intersects the radiation beams of the active nucleus (Humphrey et al. 2013), as appears to be the case in TXS 0211-122.  

In its wake, the superbubble can be expected to leave behind condensations of infalling gas, resulting from energy dissipation during collisions between individual bubbles, or from blow-outs due to mass-loading.  When irradiated by the active galactic nucleus, these condensations will emit narrow emission lines and thus may appear as an extended, infalling, ionized nebula around the galaxy (e.g., Humphrey et al. 2007; Villar-Mart\'{i}n et al. 2007; Adams et al. 2009).  

How many supernovae would be required to power the expansion of the shell?  Assuming an expansion velocity of $\sim$60 km s$^{-1}$ and gas 
mass of $\ge$4$\times$10$^8 M_{\odot}$ (see $\S$4), then the kinetic energy of the shell would be $\ga$10$^{55}$ erg.  A single
 core-collapse supernova will release 10$^{51}$ erg of kinetic energy, of which $\sim$10\% can contribute to producing and powering the shell (e.g. Thornton et al. 1998).  A total of $\ga$10$^5$ supernovae would then be required, in turn requiring the formation of $\ga$10$^7$ $M_{\odot}$ of stars 
(e.g. Heckman et al. 1990).  Although TXS 0211-122 shows no clear evidence for ongoing star formation (V01), {\it Spitzer} IRAC
 photometry does permit a star formation rate of up to $\sim$200 $M_{\odot}$ yr$^{-1}$, while also constraining the stellar mass to be
 $\le$1.4$\times$10$^{11}$  $M_{\odot}$ (Drouart et al., in preparation).  

Although it has been suggested that powerful radio jets can also sweep interstellar gas into shells (e.g. Krause 2002), 
we believe this would not provide a natural explanation in the case of TXS 0211-122.  Our main objection is that 
a spatial correlation between the radio hotspots and the limb of the shell, which would be expected if the shell is driven by the jets, is not observed.
 I.e., the Eastern radio source has an observed length of $\sim$50 kpc, and the Western one has a length of $\sim$90 kpc (Carilli et al. 1997), both of
 which are very different to the 61 kpc radius of the shell.

\section{Conclusions}
We have analyzed deep spectropolarimetry and spectroscopy of the z$=$2.34 radio galaxy
TXS 0211-122 and its gaseous environment, aiming to 
understand the origin, geometry and power source of high redshift
Ly$\alpha$ nebulae.  

Although the polarization of Ly$\alpha$ is low centrally (P$<$4.7 \%), it rises to P$=$16.4$\pm$4.6 \% at a 
distance of $\ge$12 kpc (1.5\arcsec) East of the galactic nucleus.  This indicates that the giant Ly$\alpha$
nebula is partly powered by the scattering of 
centrally-produced Ly$\alpha$ photons in a spatially extended gaseous envelope.  This is the first detection of polarized 
Ly$\alpha$ emission, to date, from a nebula around a radio loud active galaxy, and only the second from any extragalactic Ly$\alpha$ nebula. 

At the extreme edges of the Ly$\alpha$ nebula, at a radius
corresponding to 61 kpc, we have detected a pair of diametrically
opposed UV sources.  We have suggested that they are the limb of a dusty, gaseous shell
surrounding the Ly$\alpha$ nebula, seen in illuminated thanks to the presence of the powerful 
active nucleus in TXS 0211-122.  

We have argued that the observed configuration of the gaseous environment of TXS 0211-122 
can be naturally understood in terms of the expansion of a starburst driven bubble through the 
galaxy and its gaseous halo, spreading metals and dust out to radii of $\sim$61 kpc.

\section*{Acknowledgements}
AH acknowledges a Marie Curie Fellowship cofunded by the 7$^{th}$ Research Framework Programme and the Portuguese Funda\c{c}\~ao para a Ci{\^e}ncia e a Tecnologia.  Some of the data presented herein were obtained at the W.M. Keck Observatory, which is operated as a scientific partnership among the California Institute of Technology, the University of California and the National Aeronautics and Space Administration. The Observatory was made possible by the generous financial support of the W.M. Keck Foundation.  We also thank the referee for useful comments.

\setlength{\bibhang}{2.0em}

\label{lastpage}
\end{document}